Влияние инжекционной способности анодного эмиттера на характеристики комбинированных СИТ-МОП-транзисторов.

*А. С.Кюрегян*[1], *А. В. Горбатюк*[2], *Б. В. Иванов*[3],

[1]Всероссийский электротехнический институт им. В.И. Ленина.
[2]Физико-технический институт им. А.Ф. Иоффе Российской академии наук.
[3]Санкт-Петербургский государственный электротехнический университет «ЛЭТИ» им. В.И. Ульянова (Ленина).

С помощью двумерного численного моделирования изучена возможность оптимизации высоковольтных комбинированных СИТ-МОП-транзисторов (КСМТ) путем локального уменьшения времени жизни вблизи анодного эмиттера и/или снижения его инжекционной способности тремя различными способами. Показано, что с физической точки зрения все четыре способа оптимизации эквивалентны и позволяют уменьшить энергию $E_{\mathrm{off}}$ потерь в КСМТ при выключении на 30-40%, как и в биполярных транзисторах с изолированным затвором (БТИЗ). Однако при прочих равных условиях энергия $E_{\mathrm{off}}$ в КСМТ оказалась на 15-35% меньше, чем в эквивалентных БТИЗ траншейной конструкции.

## 1. Введение

В настоящее время биполярные транзисторы с изолированным затвором (БТИЗ) являются наиболее эффективными и распространенным ключевыми элементами для преобразователей средней мощности [1,2]. Наилучшими характеристиками обладают БТИЗ траншейной конструкции, содержащие дополнительный стоп-слой между $p^+$-коллектором и высокоомной $n$-базой [3] (в англоязычной литературе такой прибор называют Carrier Storage Trench Bipolar Transistor, или CSTBT), однако для его производства необходима технологическая база очень высокого уровня (см. Рис. 1a). Гораздо более прост в изготовлении комбинированный СИТ-МОП-транзистор (КСМТ), который является полным функциональным аналогом CSTBT. Это гибридный прибор, содержащий высоковольтный тиристор с электростатическим управлением (СИТ) и управляющий низковольтный МОП-транзистор, соединенный с ним по каскодной схеме [4] (см. Рис. 1b). Результаты экспериментальных исследований опытных образцов КСМТ [5,6] и численного моделированиям [7,8] показали, что эти приборы превосходят CSTBT по ряду параметров. Однако в [5-8] не изучались возможности минимизации потерь в КСМТ путем локального уменьшения времени жизни вблизи анодного эмиттера или снижения его инжекционной способности, которые были предложены для

БТИЗ [9-12] и применимы для любых биполярных переключателей с распределенными по площади микрозатворами [13]. В настоящей работе мы изучим возможность оптимизации СИТ подобным образом с помощью численного моделирования. Будет показано, что (а) с физической точки зрения все способы эквивалентны, (б) оптимизация позволяет уменьшить энергию потерь в КСМТ при выключении $E_{\rm off}$ на 30-40% при фиксированном падении напряжения на открытом состоянии $U_{\rm on}$ и (в) при прочих равных условиях энергия потерь $E_{\rm off}$ в КСМТ во всех случаях на 15-35% меньше, чем в CSTBT.

## 2. Конструкции приборов и метод моделирования их характеристик.

Значения геометрических и электрофизических параметров КСМТ и CSTBT, для которых выполнялись расчеты, приведены в Таблице 1. Распределения легирующих примесей в истоке и затворе СИТ, а также в истоке и коллекторе БТИЗ задавалось функциями Гаусса. Легирование стоп-слоя $n_1$ в CSTBT предполагалось однородным. Распределения акцепторов в анодном эмиттере и доноров в буферном $n'$-слое задавалось функцией ошибок (при $h_p = 10$ мкм и $h_n = 20$ мкм) или функцией Гаусса (при $h_p = 2$ мкм и $h_n = 12$ мкм).

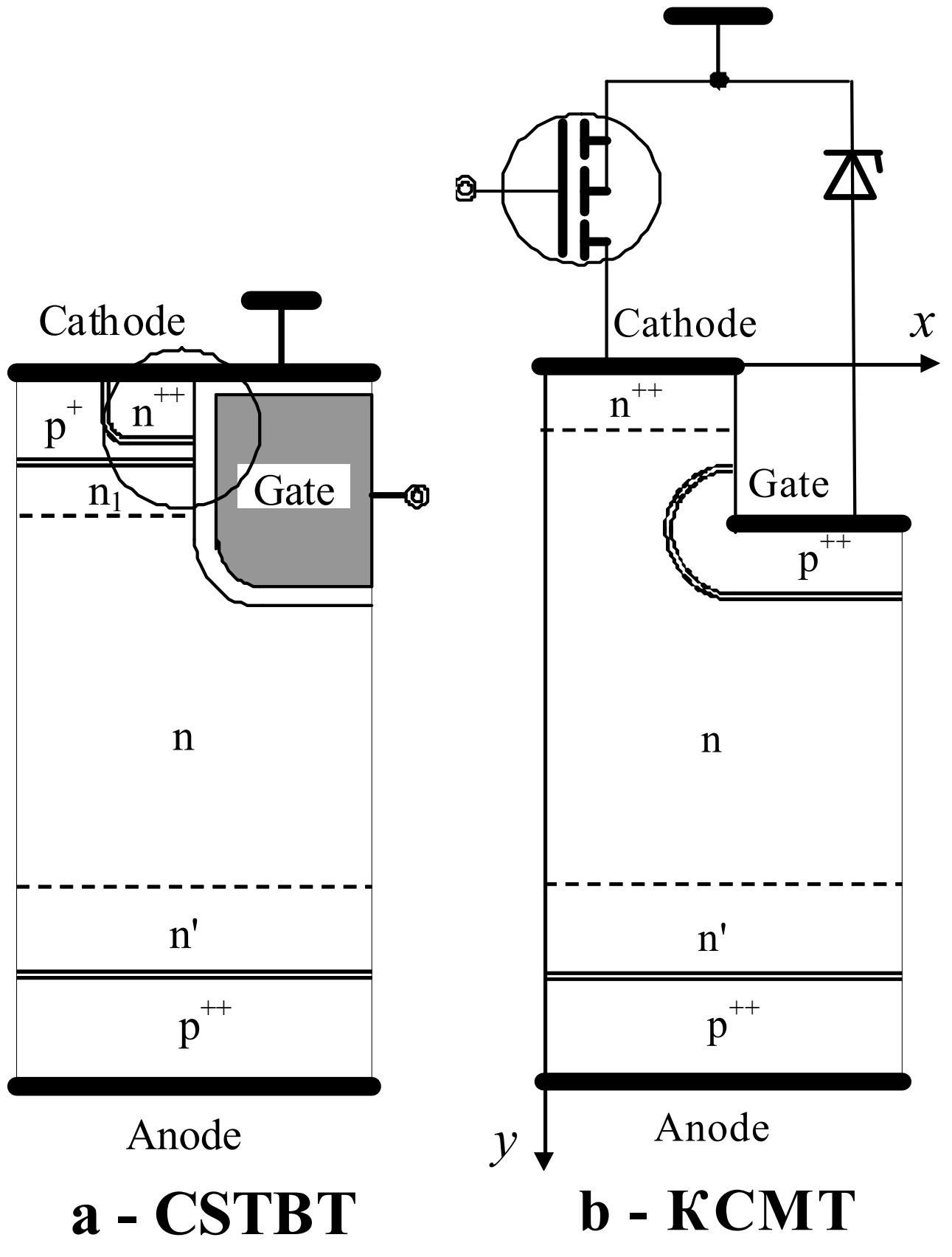

Рис. 1. Схематичное изображение поперечной ячейки CSTBT и комбинированного СИТ-МОП-транзистора (КСМТ). Окружностью выделен интегрированный в основной чип CSTBT МОП-транзистор, который в КСМТ изготовливается на отдельном чипе.

**Таблица 1.**

| Наименование параметров | Обозначения | Значения параметров |
|---|---|---|
| **Параметры управляемого катода СИТ.** | | |
| Глубина диффузии доноров в истоке | $x_s$, мкм | 1.5 мкм |
| Поверхностная концентрация доноров в истоке | $N_s$ | $10^{20}$ см$^{-3}$ |
| Глубина *p-n*–перехода затвора | $x_g$ | 3 мкм |
| Поверхностная концентрация акцепторов в затворе | $N_g$ | $10^{18}$ см$^{-3}$ |
| **Параметры управляемого катода БТИЗ**. | | |
| Толщина подзатворного диэлектрика | $h_{\mathrm{SiO}}$ | 0,015 мкм |
| Ширина $n^+$-истока | $l_s$ | 1,67 мкм |
| Глубина перехода $n^+$-истока | $x_s$ | 1,5 мкм |
| Поверхностная концентрация доноров в $n^+$-истоке | $N_s$ | $10^{20}$ см$^{-3}$ |
| Глубины перехода коллектора | $x_c$ | 4 мкм |
| Поверхностная концентрация акцепторов в коллекторе | $N_c$ | $10^{18}$ см$^{-3}$ |
| Толщина $n_1$-стоп-слоя | $h_1$ | 1,5 мкм |
| Концентрация доноров в $n_1$-стоп-слое | $N_1$ | $10^{16}$ см$^{-3}$ |
| **Общие постоянные параметры БТИЗ и СИТ** | | |
| Ширина ячейки | $L$ | 10 мкм |
| Глубина затворной траншеи | $h$ | 8 мкм |
| Ширина затворной траншеи | $l_g$ | 5 мкм |
| Толщина высокоомной *n*-базы | $w$ | 450 мкм |
| Концентрация доноров в *n*-базе | $N_d$ | $1{,}75\cdot10^{13}$ см$^{-3}$ |
| Напряжение пробоя | $U_b$ | 4,6 кВ |
| **Общие переменные параметры БТИЗ и СИТ** | | |
| Время жизни электронов и дырок в *n*-базе | $\tau_0$ | 2-64 мкс |
| Время жизни электронов и дырок в *n'*-слое | $\tau_{\mathrm{buff}}$ | 0,03-64 мкс |
| Поверхностная концентрация доноров в буферном *n'*-слое | $N_{\mathrm{buff}}$ | $7\cdot10^{16}$-$4\cdot10^{18}$ см$^{-3}$ |
| Толщина буферного *n'*-слоя | $h_n$ | 20 мкм или 12 мкм |
| Поверхностная концентрация акцепторов в анодном $p^+$-слое | $N_{\mathrm{A}}$ | $1{,}8\cdot10^{18}$-$10^{20}$ см$^{-3}$ |
| Толщина анодного $p^+$-слоя | $h_p$ | 10 мкм или 2 мкм |

Как известно, электронные процессы в подобных структурах описываются системой двух уравнений непрерывности и уравнения Пуассона, в которых должны быть учтены все нелинейные эффекты, в том числе, Оже-рекомбинация, зависимости подвижностей $\mu_{e,h}$, коэффициентов диффузии $D_{e,h}$, ширины запрещенной зоны $\mathcal{E}_g$ и времени жизни $\tau$ от суммарной концентрации $N$ легирующих примесей. Проблема моделирования усугубляется тем, что структура СИТ и БТИЗ сильно неоднородна по объему, а процессы в прикатодных областях существенно неодномерны. С целью адекватного численного решения столь сложной задачи мы использовали программный продукт TCAD SENTAURUS фирмы [14]. Для построения неоднородной сетки конечных элементов, соответствующей структурам СИТ и БТИЗ, использовались стандартные процедуры этой программы, обеспечивающие «сгуще-

ние» сетки в областях с наибольшими градиентами концентрации легирующих примесей и, особенно, в инверсионном канале МОП-транзистора, встроенного в БТИЗ. Зависимости $\mu_{e,h}(N)$, $D_{e,h}(N)$ и $\mathcal{E}_g(N)$ описывались формулами, предложенными в работах [15] и [16] соответственно. Для определенности мы считали равными параметры $\tau_{n0}, \tau_{p0}$ в формуле Шокли−Рида для рекомбинации через глубокий уровень, расположенный в середине запрещенной зоны кремния. Зависимость $\tau(N)$ описывалась формулой

$$\tau_{p0}^{-1} = \tau_{n0}^{-1} = \tau^{-1} = \tau_0^{-1} + \tau_N^{-1}(1 + N/N_c), \tag{1}$$

где $\tau_0$ время жизни, соответствующее рекомбинации через однородно распределенные центры, введение которых (например, с помощью электронного или γ-облучения) позволяет регулировать быстродействие приборов, а второе слагаемое в правой части описывает увеличение концентрации «естественных» центров рекомбинации с ростом уровня легирования донорами и акцепторами. Мы использовали значения параметров $\tau_N = 4$ мкс и $N_c = 10^{17}$ см$^{-3}$.

В использованной нами модели коммутация приборов осуществлялась путем мгновенного изменения потенциалов затворов управляющих МОП-транзисторов с -20 В до +20 В (при включении) и обратно (при выключении). Согласно расчетам [7] при этом приведенное сопротивление $R_{\mathrm{dsON}}$ открытого МОП-транзистора в CSTBT равно примерно 1 мОм·см$^2$. Это же значение $R_{\mathrm{dsON}}$ мы использовали для управляющего МОП-транзистора в КСМТ. В качестве регулирующего элемента КСМТ, соединяющого затвор СИТ с заземленным электродом КСМТ, мы использовали стабистор [4] с напряжением стабилизации 1,6 В и емкостью при нулевом смещении 20 нФ/см$^2$.

### 3. Результаты моделирования и их обсуждение.

В этом разделе приведены результаты моделирования коммутационных характеристик приборов в цепи с активной нагрузкой 24 Ом·см$^2$ , напряжением источника питания $U_{\mathrm{off}} = 2.4$ кВ и плотностью тока в открытом состоянии $J_{\mathrm{on}} = 100$ А/см$^2$. Типичная зависимость мощности $P = U_A J_A$ ($U_A$ и $J_A$ - мгновенные значения анодного напряжения и плотности тока), рассеиваемой КСМТ, от времени приведена на Рис. 2. Средняя по времени плотность мощности потерь равна

$$P_{av} = \left(U_{\mathrm{on}} J_{\mathrm{on}} T_{\mathrm{on}} + E_{\mathrm{off}} + E_{\mathrm{on}}\right) T^{-1}, \tag{2}$$

где $T_{\mathrm{on}}$ - длительность импульсов, $T^{-1}$ - частота их повторения, а

$$E_{\mathrm{on}} = \int_0^{T_{\mathrm{on}}} \left(P - U_{\mathrm{on}} J_{\mathrm{on}}\right) dt, \quad E_{\mathrm{off}} = \int_{T_{\mathrm{on}}}^{T} P dt \tag{3}$$

- плотности энергии коммутационных потерь при включении и выключении. Оптимизация приборов состоит в выборе конструкции, которая обеспечит минимальную мощность $P_{av}$ при заданных значениях $U_b, J_{\text{on}}, U_{\text{off}}, T_{\text{on}}$. В стандартных приборах с высокоэффективным анодным эмиттером для этого используется электронное или γ-облучение, которое уменьшает время жизни $\tau_0$ однородно по всей толщине структуры. В дополнение к этому мы использовали четыре других способа:

1) локальное уменьшение времени жизни $\tau_0$ до величины $\tau_{\text{buff}}$ в прианодной области буферного слоя толщиной 6 мкм при $N_{\text{buff}} = 7\cdot10^{16}$ см$^{-3}$, $h_n = 20$ мкм, $N_{\text{A}} = 10^{20}$ см$^{-3}$, $h_p = 10$ мкм; распределения легирующих примесей и времени жизни вблизи анода изображены для этого случая на Рис. 3.

2) уменьшение толщины $h_p$ анодного эмиттера до 2 мкм и изменение поверхностной концентрации акцепторов $N_{\text{A}}$ в пределах $(0.07-2)\cdot10^{18}$ см$^{-3}$ при $\tau_{\text{buff}} = \tau_0$, $N_{\text{buff}} = 7\cdot10^{16}$ см$^{-3}$ и $h_n = 20$ мкм;

3) изменение поверхностной концентрации доноров $N_{\text{buff}}$ в пределах $(0.02-1)\cdot10^{20}$ см$^{-3}$ при $h_n = 12$ мкм, $N_{\text{A}} = 10^{20}$ см$^{-3}$, $h_p = 2$ мкм и $\tau_{\text{buff}} = \tau_0$;

4) изменение поверхностной концентрации доноров $N_{\text{buff}}$ в пределах $(0.6-5)\cdot10^{17}$ см$^{-3}$ при $h_n = 20$ мкм, $N_{\text{A}} = 10^{20}$ см$^{-3}$, $h_p = 10$ мкм и $\tau_{\text{buff}} = \tau_0$.

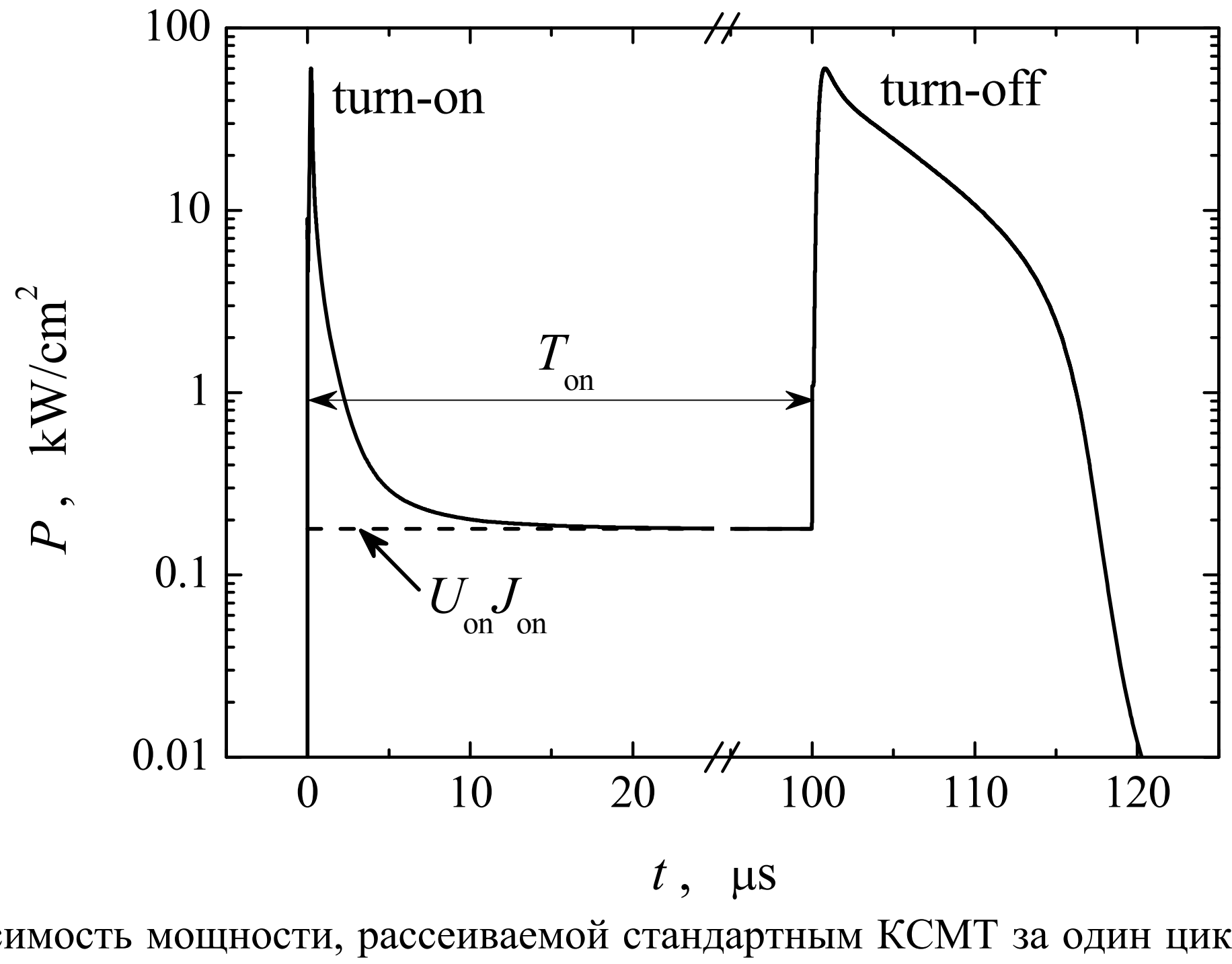

Рис. 2. Зависимость мощности, рассеиваемой стандартным КСМТ за один цикл коммутации при $\tau_0 = \tau_{\text{buff}} = 4$ мкс, $N_{\text{buff}} = 7\cdot10^{16}$ см$^{-3}$ и $N_{\text{A}} = 10^{20}$ см$^{-3}$.

В последних двух случаях увеличение $N_{\text{buff}}$ приводило не только к ухудшению инжекционной способности анодного эмиттера, но и к уменьшению времени жизни в буферном слое в соответствии с формулой (1). Поэтому способы 3 и 4 являются фактически комбинациями способов 1 и 2.

Величины $U_{\text{on}}$ и $E_{\text{off}}$ определяются главным образом распределениями $p_{\text{on}}(y) \approx n_{\text{on}}(y)$ неравновесных дырок и электронов в толстой высокоомной *n*-базе открытых приборов. Так как вдали от катода (при $y > 3L$) эти распределения практически одномерны (то есть не зависят от *x*) [17], то влияние на них электрофизических и геометрических характеристик буферного и анодного слоев проявляется через единственный параметр - отношение $\gamma_{pB} = J_p / J$ плотности тока дырок $J_p$ на границе между *n*-базой и буферным слоем (в нашем случае при $y = 450$ мкм – см. рис. 3) к полной плотности тока анода $J$. По сути дела $\gamma_{pB}$ является коэффициентом инжекции дырок виртуальным эмиттером, состоящим из буферного слоя и реального анодного $p^+$-эмиттера. Величина $\gamma_{pB}$ зависит от $\tau_{\text{buff}}$, профиля легирования буферного слоя и соотношения между концентрациями $N_{\text{A}}$ и $N_{\text{buff}}$. Поэтому все четыре описанных выше способа позволяют изменять $\gamma_{pB}$, как это изображено на Рис. 4. Уменьшение $\gamma_{pB}$ приводит к изменениям распределений $p_{\text{on}}(y)$, примеры которых приведены на Рис. 5.

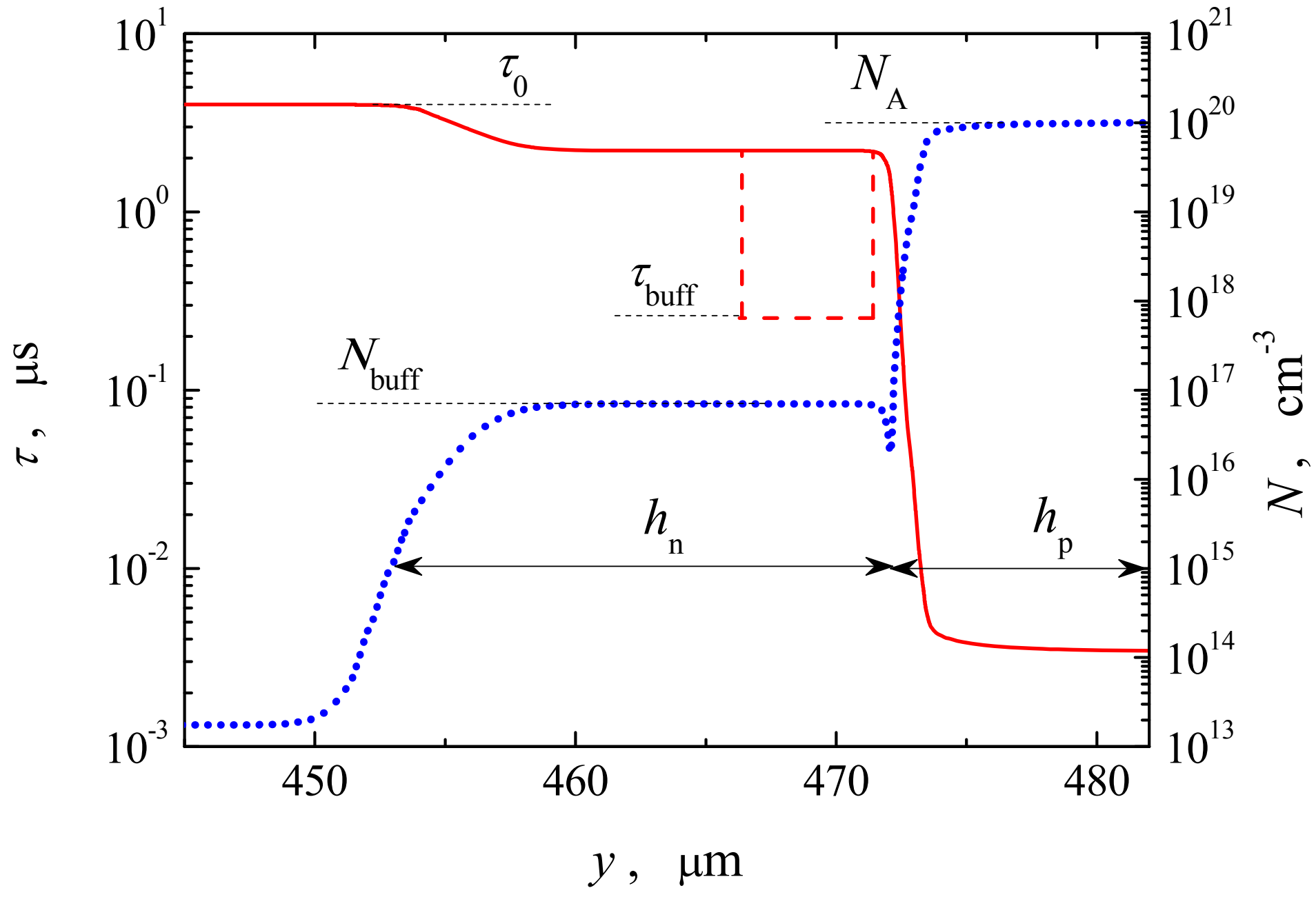

Рис. 3. Распределения легирующих примесей $N$ (точки) и времени жизни носителей заряда $\tau$ (линии) около анода Вертикальные штриховые линии ограничивают область, в которой время жизни уменьшается до величины $\tau_{\text{buff}} \le \tau$.

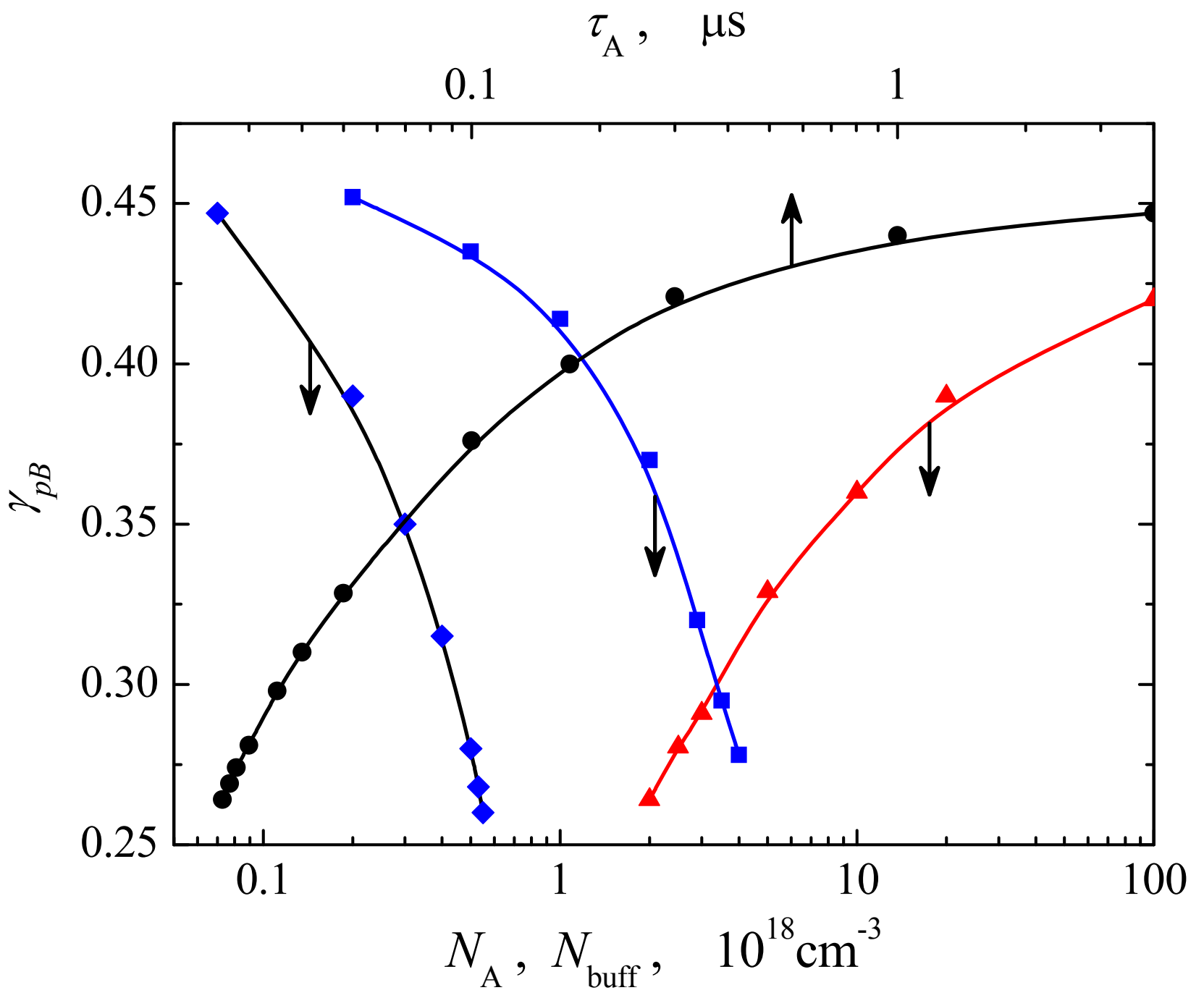

Рис. 4. Зависимости коэффициента инжекции дырок $\gamma_{pB}$ в $n$-базу открытого СИТ с $\tau_0 = 4$ мкс от $\tau_{buff}$ (кружки, способ 1), от $N_A$ (треугольники, способ 2) и от $N_{buff}$ (квадраты - способ 3 и ромбы – способ 4).

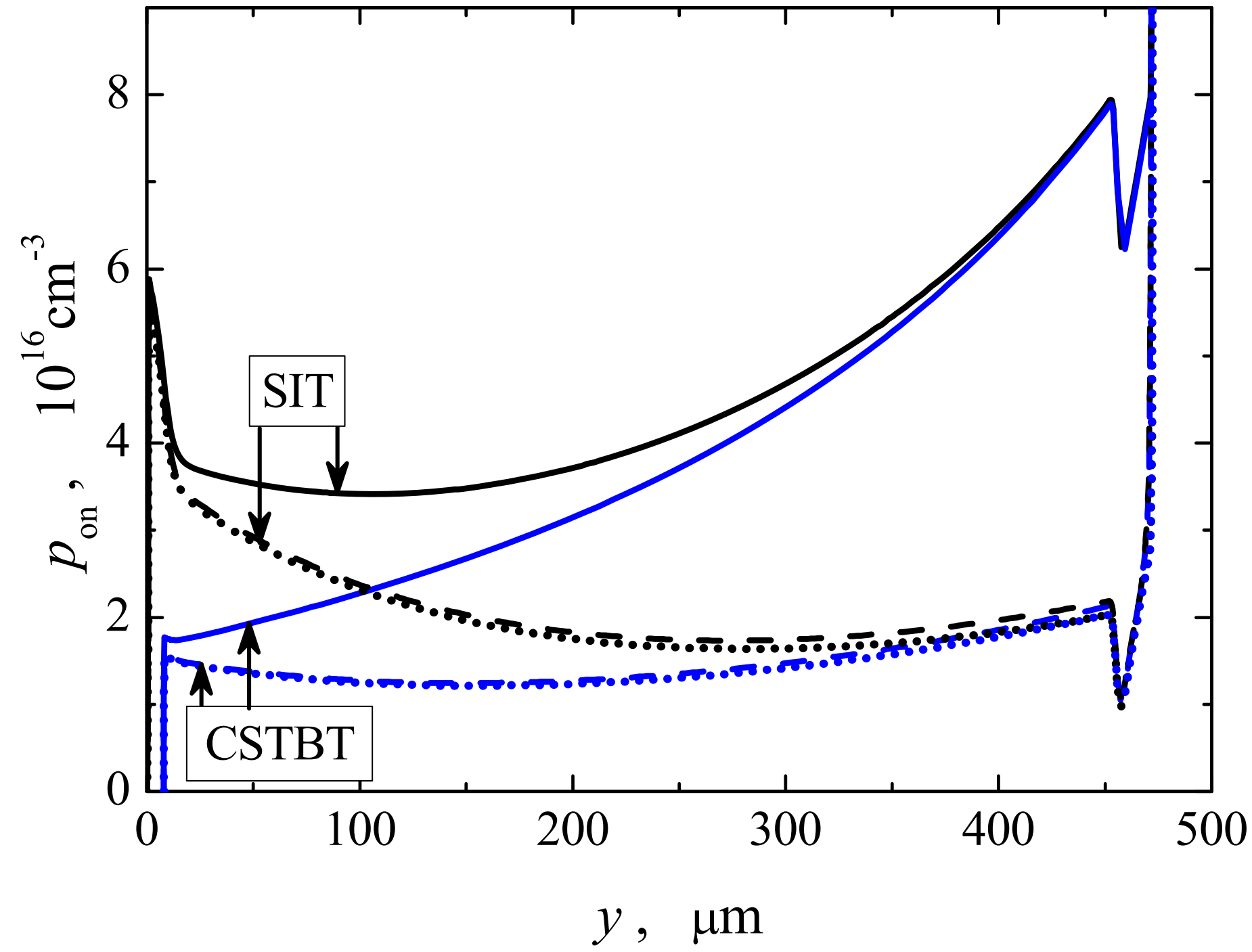

Рис. 5. Распределения дырок в открытых СИТ и БТИЗ при $\tau_0 = 16$ мкс, $N_{buff} = 7 \cdot 10^{16}$ см$^{-3}$ и $\tau_{buff} = 16$ мкс, $N_A = 10^{20}$ см$^{-3}$ (сплошные линии, стандартные приборы), $\tau_{buff} = 35$ нс, $N_A = 10^{20}$ см$^{-3}$ (штриховые линии, способ 1), $\tau_{buff} = 16$ мкс, $N_A = 3.5 \cdot 10^{18}$ см$^{-3}$ (точки, способ 2).

Как и следовало ожидать, при уменьшении $\gamma_{pB}$ падение напряжения $U_{\text{on}}$ в открытом состоянии и энергия $E_{\text{on}}$, рассеиваемая при включении, увеличиваются, а энергия $E_{\text{off}}$, рассеиваемая при выключении, уменьшается. Причина таких изменений, пример которых приведен на Рис. 6, очевидна: уменьшение $\gamma_{pB}$ приводит к замедлению процесса наполнения $n$-базы дырками (и, следовательно, к росту $E_{\text{on}}$), уменьшению концентрации плазмы в открытом состоянии (то есть к снижению проводимости $n$-базы и росту $U_{\text{on}}$) и уменьшению времени извлечения накопленного в базе заряда при выключении (то есть к снижению $E_{\text{off}}$). Так как способ регулирования величины $\gamma_{pB}$ практически не влияет на изменение распределений $p_{\text{on}}(y)$ (см. Рис. 5), то зависимости $U_{\text{on}}(\gamma_{pB})$ и $E_{\text{off}}(\gamma_{pB})$ для всех способов совпадают (см. Рис. 7). Однако энергия $E_{\text{on}}$ зависит не только от $\gamma_{pB}$, но и от распределения доноров $N(y)$ в буферном слое, как это изображено на Рис. 8. Если $N(y)$ не изменяется (как при использовании способов 1 и 2), то зависимости $E_{\text{on}}(\gamma_{pB})$ совпадают. При использовании способа 3 время пролета дырок через буферный слой $t_B$ уменьшается из-за уменьшения $h_n$ и появления встроенного поля, поэтому процесс наполнения $n$-базы дырками при включении ускоряется, а энергия $E_{\text{on}}$ уменьшается при тех же значениях $\gamma_{pB}$. Напротив, при использовании способа 4 подвижность дырок в буферном слое уменьшается из за роста $N_{buff}$, поэтому время $t_B$ и энергия $E_{\text{on}}$ увеличиваются. Заметим также, что величина $E_{\text{on}}$ практически не зависит от времени жизни $\tau_0$ в $n$-базе при прочих равных условиях.

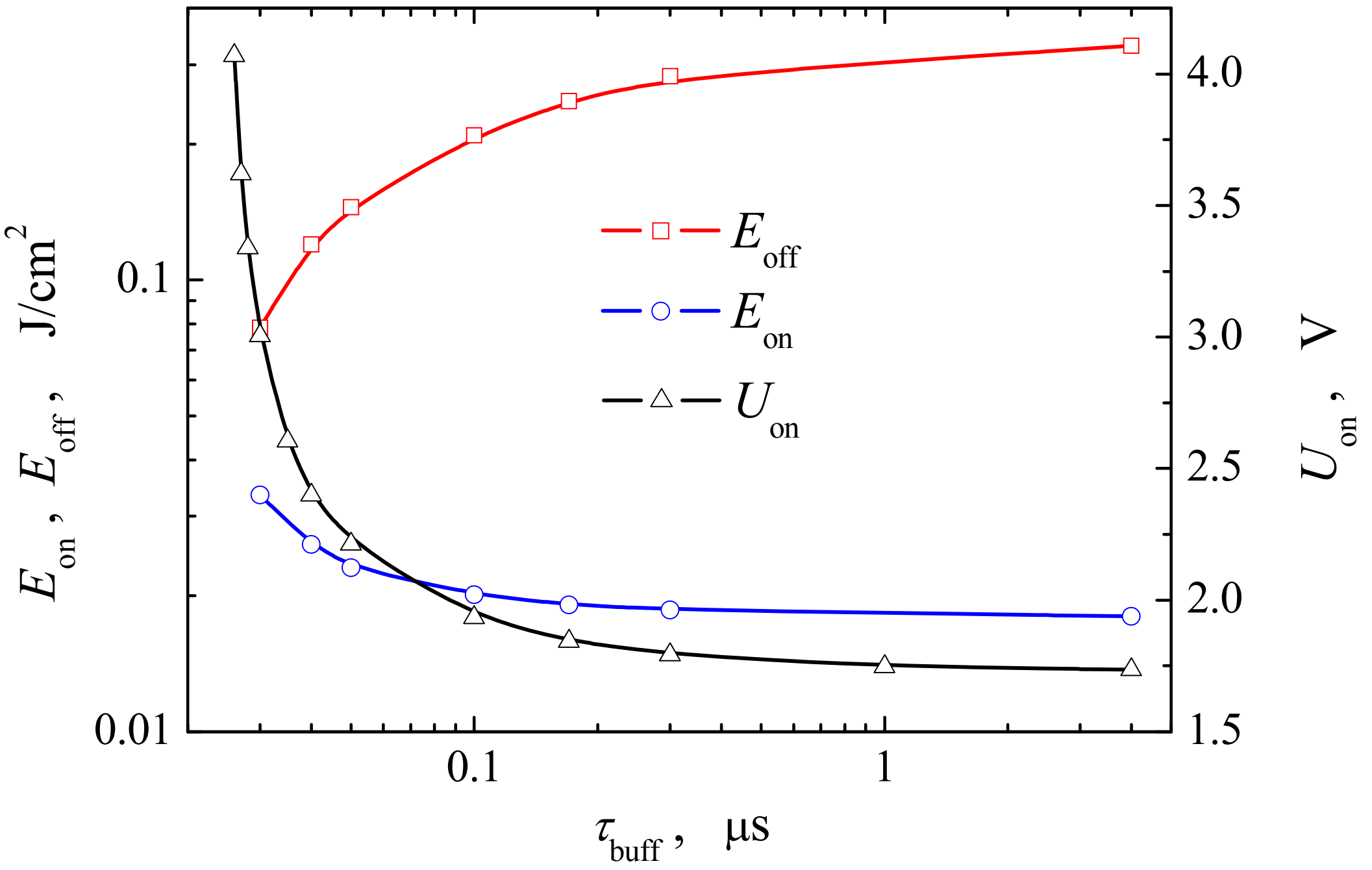

Рис. 6. Зависимости энергий потерь в КСМТ при выключении $E_{\text{off}}$ (квадраты), включении $E_{\text{on}}$ (кружки) и напряжения в открытом состоянии $U_{\text{on}}$ (треугольники) от $\tau_{\text{buff}}$ (способ 1) при $\tau_0 = 4$ мкс.

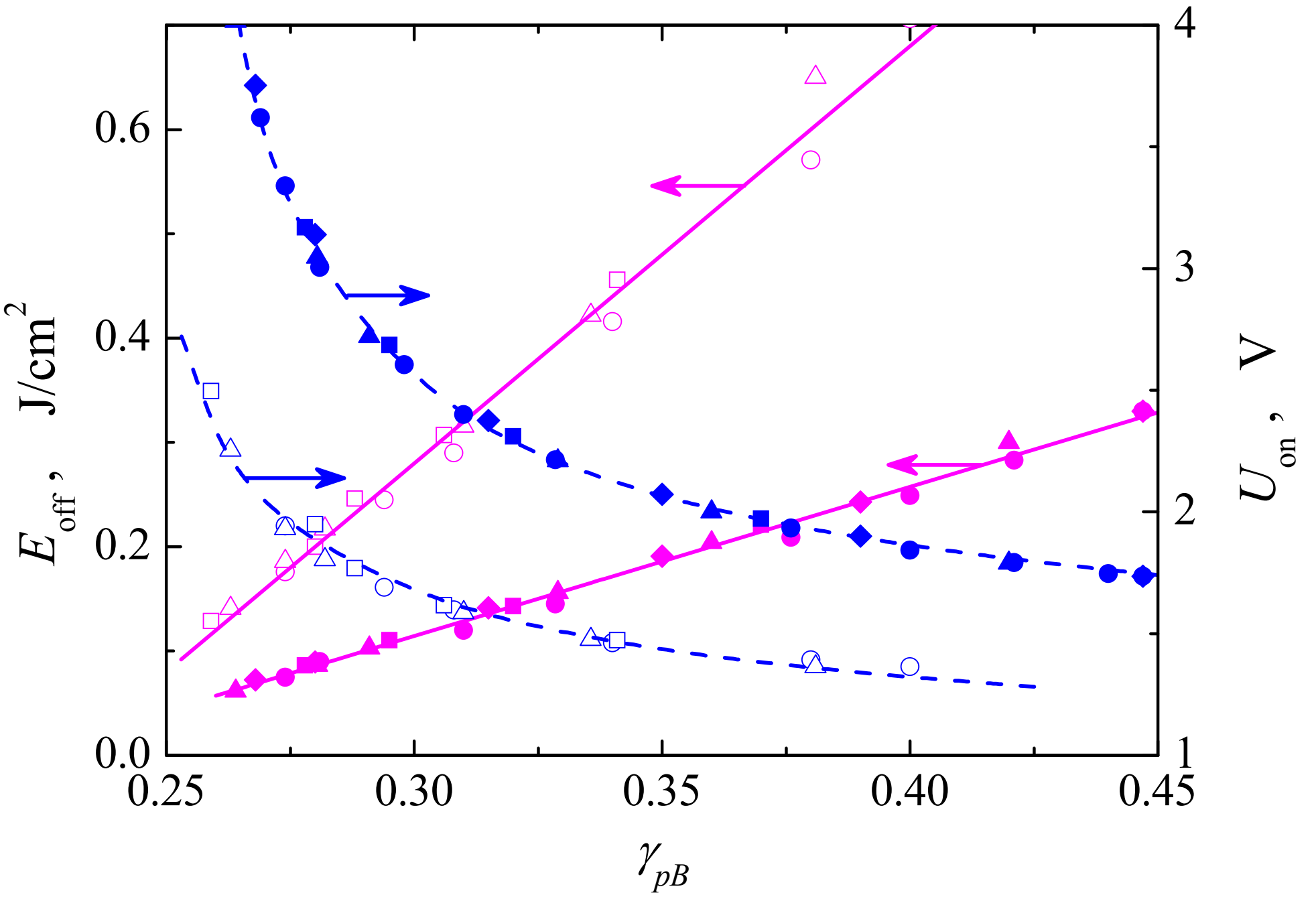

Рис. 7. Зависимости энергии потерь КСМТ при выключении $E_{\text{off}}$ (сплошные линии) и напряжения в открытом состоянии $U_{\text{on}}$ (штриховые линии) при $\tau_0 = 4$ мкс (темные символы) и $\tau_0 = 16$ мкс (светлые символы) от коэффициента инжекции $\gamma_{pB}$, который изменяется при варьировании $\tau_{\text{buff}}$ (кружки, способ 1), $N_{\text{A}}$ (треугольники, способ 2) и $N_{buff}$ (квадраты - способ 3 и ромбы – способ 4).

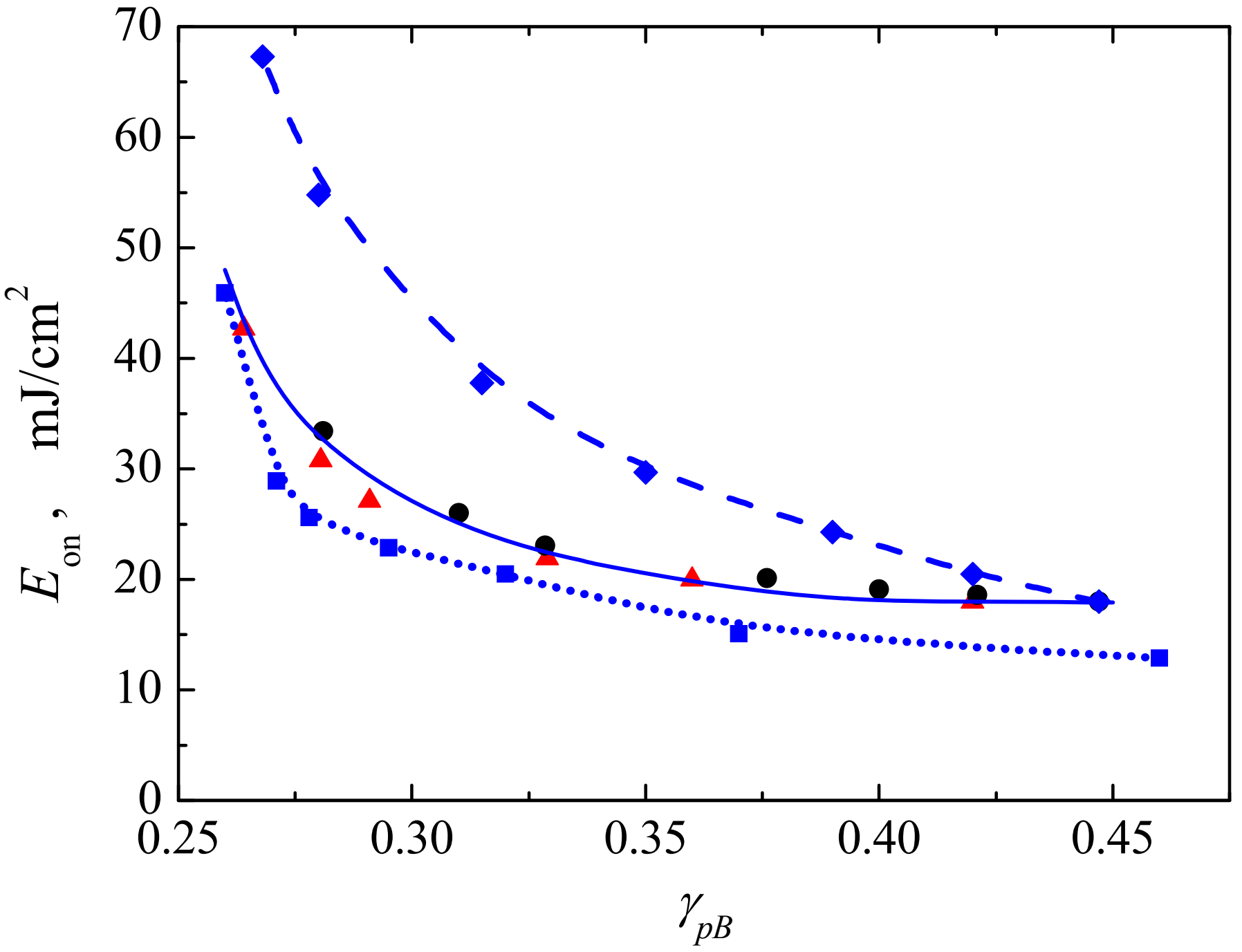

Рис. 8. Зависимости энергии потерь КСМТ при включении $E_{\text{on}}$ при $\tau_0 = 4$ мкс от коэффициента инжекции $\gamma_{pB}$, который изменяется при варьировании $\tau_{\text{buff}}$ (кружки, способ 1), от $N_{\text{A}}$ (треугольники, способ 2) и от $N_{buff}$ (квадраты - способ 3 и ромбы – способ 4).

В стандартных приборах с большими $\gamma_{pB}$ обычно выполняется сильное неравенство $E_{\text{off}} >> E_{\text{on}}$. Поэтому для минимизации $P_{\text{av}}$ используются так называемые «оптимизационные кривые» (trade-off curves), определяющие взаимосвязь между $E_{\text{off}}$ и $U_{\text{on}}$ при рабочих плотности тока и коммутируемом напряжении. Совпадение зависимостей $U_{\text{on}}\left(\gamma_{pB}\right)$ и $E_{\text{off}}\left(\gamma_{pB}\right)$ для всех способов регулирования $\gamma_{pB}$ приводит к тому, что так кривые $E_{\text{off}}\left(U_{\text{on}}\right)$ также совпадают (см. Рис. 9). Однако, в отличие от стандартного способа повышения быстродействия приборов путем уменьшения $\tau_0$, уменьшение $\gamma_{pB}$ приводит не только к снижению $E_{\text{off}}$, но и к росту $E_{\text{on}}$, особенно сильному при малых $\gamma_{pB}$. Кроме того, при приближении $\gamma_{pB}$ к некоторому критическому значению (примерно равному 0,26 в рассмотренных нами случаях) напряжение $U_{\text{on}}$ начинает очень резко увеличиваться и достигает сотен и даже тысяч вольт, так что фактически прибор перестает включаться. Это накладывает очень жесткие требования на точность воспроизведения параметров буферного слоя и анодного эмиттера при $\gamma_{pB} < 0.3$.

Для оптимизации приборов с использованием формулы (2) можно использовать аппроксимации

$$E_{\text{on}} = \hat{E}_{\text{on}} \exp\left(U_{\text{on}} / \hat{U}_{\text{on}}\right), \quad E_{\text{off}} = \frac{\hat{E}_{\text{off}}}{\left(U_{\text{on}} / \hat{U}_{\text{off}} - 1\right)^{a}}, \tag{4}$$

где $\hat{E}_{\text{on,off}}$, $\hat{U}_{\text{on,off}}$ и $a$ - подгоночные параметры, которые зависят от $U_b, J_{\text{on}}, U_{\text{off}}$ и $\tau_0$. В рассмотренных нами случаях наилучшее совпадение с результатами моделирования получилось при $\hat{E}_{\text{on}} = 18.5$ мДж/см$^2$, $\hat{U}_{\text{on}} = \infty$, $\hat{E}_{\text{off}} = 0.2$ Дж/см$^2$, $\hat{U}_{\text{off}} = 1.15$ В, $a = 0.73$ для стандартного прибора с постоянным значением $\gamma_{pB} = 0.45$, $\hat{E}_{\text{on}} = 4.5$ мДж/см$^2$, $\hat{U}_{\text{on}} = 1$ В, $\hat{E}_{\text{off}} = 0.12$ Дж/см$^2$, $\hat{U}_{\text{off}} = 1.23$ В, $a = 0.75$ при $\tau_0 = 16$ мкс и $\hat{E}_{\text{on}} = 9.3$ мДж/см$^2$, $\hat{U}_{\text{on}} = 2.5$ В, $\hat{E}_{\text{off}} = 0.09$ Дж/см$^2$, $\hat{U}_{\text{off}} = 1.5$ В, $a = 0.7$ при $\tau_0 = 4$ мкс для приборов с пониженными значениями $\gamma_{pB}$.

Все описанные выше закономерности в равной степени относятся и к изученным нами CSTBT. Количественные различия обусловлены относительно низкой инжекционной способностью катодного эмиттера CSTBT. Вследствие этого уменьшение $\gamma_{pB}$ приводит к тому (см. Рис. 5), что распределение $p_{\text{on}}\left(y\right)$ в CSTBT превращаются из наихудшего (когда максимум $p_{\text{on}}$ расположен вблизи анода - вариант **A** по терминологии работы [9]) в промежуточное (когда $p_{\text{on}} \approx \text{const}$ - вариант **B**), а в СИТ – из наихудшего в оптимальное (когда максимум $p_{\text{on}}$ расположен вблизи катода - вариант **C**). Сравнение «оптимизационных кривых», приве-денных на Рис. 10, показывает, что при одинаковых значениях $U_{on}$ и прочих равных условиях энергия $E_{off}$ потерь в СИТ оказывается заметно меньше (на 15-35% в рассмотренных нами случаях), чем в эквивалентных CSTBT.

Работа выполнена при поддержке РФФИ (грант 13-07-00943).

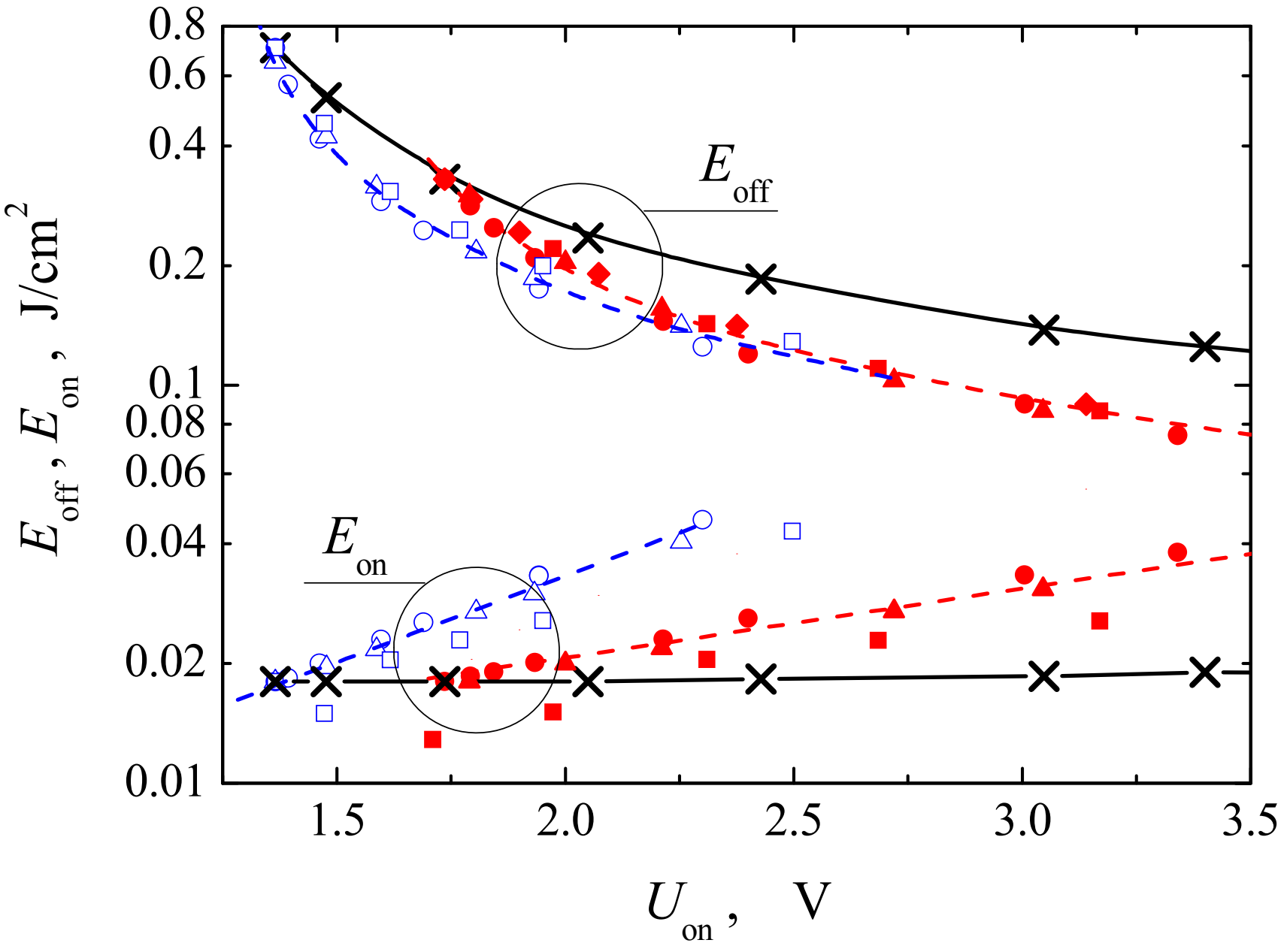

Рис. 9. Взаимосвязи между $E_{off}$ и $U_{on}$ при изменении $\tau_0 = \tau_{buff} = 2 - 64$ мкс ($N_{buff} = 7 \cdot 10^{16}$ см$^{-3}$, $N_A = 10^{20}$ см$^{-3}$, кресты), $\tau_{buff}$ (способ 1 -кружки), $N_A$ (способ 2 - треугольники), и $N_{buff}$ (способы 3 - квадраты и 4 - ромбы) при $\tau_0 = 16$ мкс (светлые символы), $\tau_0 = 4$ мкс (темные символы). Линии – расчет по формулам (4).

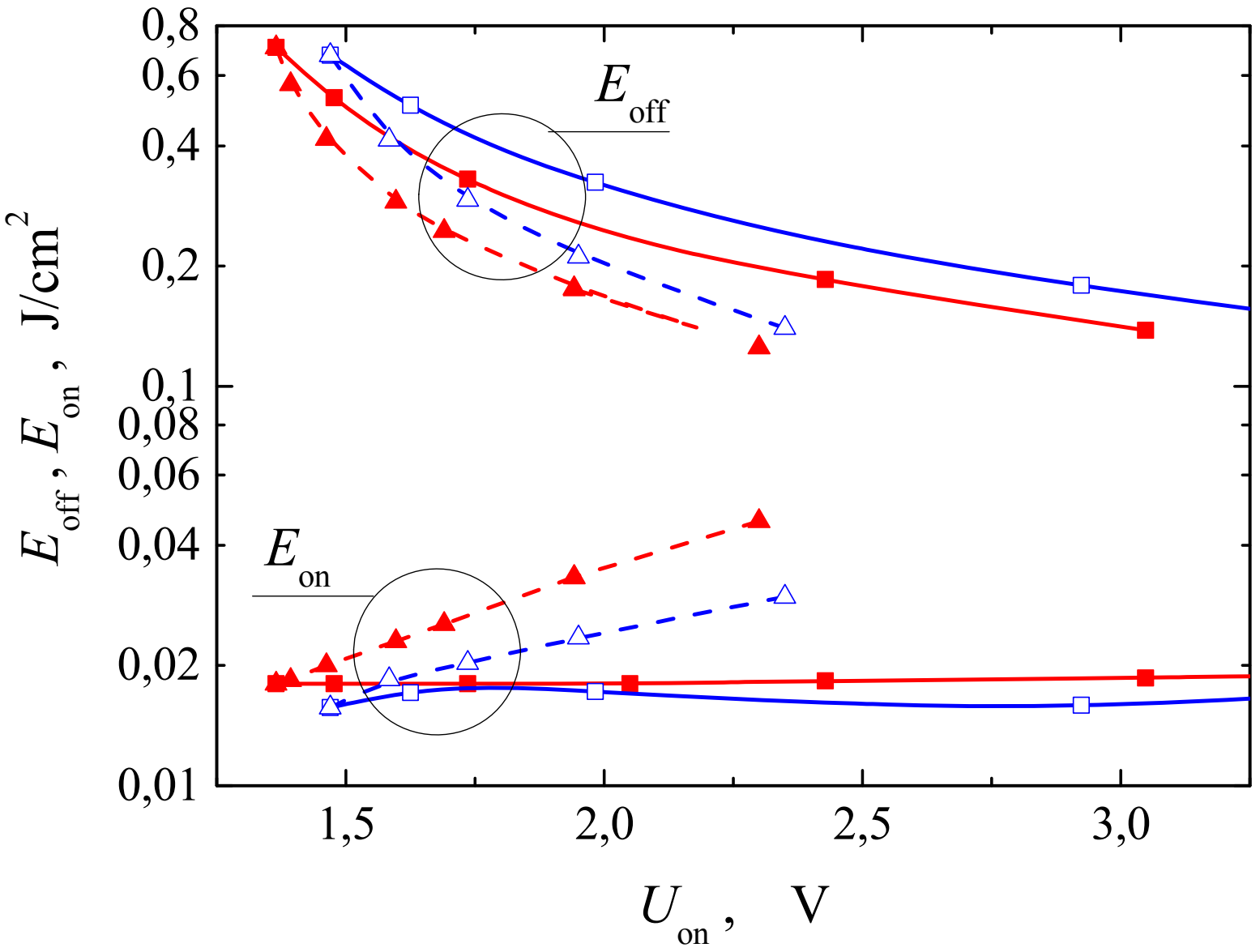

Рис. 10. Взаимосвязи между $E_{off}$ и $U_{on}$ для КСМТ (темные символы) и CSTBT (светлые символы) при изменении $\tau_0 = \tau_{buff} = 2 - 64$ мкс (квадраты) или только $\tau_{buff}$ при $\tau_0 = 16$ мкс (кружки, способ 1).

## Литература